\documentclass[amsmath,amssymb,aps,pra,reprint,superscriptaddress,footnoteinbib]{revtex4-1} %longbibliography,
\usepackage{amsmath,amssymb}
\usepackage[english]{babel}
\usepackage{ bbold }
\usepackage{upgreek}
\usepackage{dsfont}
\usepackage{graphicx}
\usepackage{comment}
\usepackage{braket}
\usepackage{colortbl}
\usepackage{blindtext}
\usepackage{hyperref}
\begin{document}
	
\title{Theory of nanoparticle cooling by elliptic coherent scattering}

\author{Henning Rudolph}
\affiliation{Faculty of Physics, University of Duisburg-Essen, Lotharstra\ss e 1, 47048 Duisburg, Germany}	

\author{Jonas Sch\"afer}
\affiliation{Faculty of Physics, University of Duisburg-Essen, Lotharstra\ss e 1, 47048 Duisburg, Germany}

\author{Benjamin A. Stickler}
\affiliation{Faculty of Physics, University of Duisburg-Essen, Lotharstra\ss e 1, 47048 Duisburg, Germany}
\affiliation{QOLS, Blackett Laboratory, Imperial College London, London SW7 2AZ, United Kingdom}

\author{Klaus Hornberger}
\affiliation{Faculty of Physics, University of Duisburg-Essen, Lotharstra\ss e 1, 47048 Duisburg, Germany}

\date{\today}	
	
\begin{abstract}
    Coherent scattering of an elliptically polarised tweezer into a cavity mode provides a promising platform for cooling levitated nanoparticles into their combined rotational and translational quantum regime [Phys. Rev. Lett. \textbf{126}, 163603 (2021)]. This article presents the theory of how aspherical nanoparticles are affected by elliptically polarised laser beams, how two orthogonal cavity modes enable rotational and translational cooling, and how the resulting power spectra contain signatures of rotational non-linearities. We provide analytic expressions for the resulting trapping frequencies, opto-mechanical coupling strengths, cooling rates, and steady-state occupations and we study their dependence on the tweezer ellipticity.
\end{abstract}
	
\maketitle
	
\section{Introduction} 
	
Preparing the mechanical motion of nanoscale dielectrics in the deep quantum regime is a longstanding goal in levitated optomechanics \cite{millen2020optomechanics}, with great potential for fundamental tests and technological applications \cite{millen2020quantum,moore2020searching}. The recent achievements of center-of-mass quantum cooling \cite{delic2020,tebbenjohanns2020} and 
rotational cooling \cite{delord2020,bang2020} are important steps towards fully controlling the 6D dynamics of microscale objects at the quantum limit. Elliptic coherent scattering cooling \cite{prlsubmission} offers as an attractive platform for cooling the combined rotational and translational motion of aspherical nanoparticles.

Their ability to rotate distinguishes levitated particles from from other optomechanical systems, such as clamped oscillators or levitated atoms \cite{aspelmeyer2014}. The non-linearity and non-harmonicity of rotational dynamics renders them attractive for quantum experiments \cite{stickler2018,ma2020}. However, the non-linear coupling between different rotational degrees of freedom  \cite{goldstein} also complicates cooling into the deep quantum regime \cite{stickler2016,zhong2017,seberson2019}. For instance, in a linearly polarized laser field the angular momentum component along the polarization axis is conserved for particles much smaller than the wavelength. The resulting motion  couples gyroscopically to the other rotational degrees of freedom, making cooling below the non-linearity inefficient \cite{seberson2019,bang2020}. Using elliptically polarized laser traps breaks the symmetry and therefore holds the potential of simultaneously cooling rotations and translations to the quantum limit.

Circularly and elliptically polarised lasers exert a conservative optical torque as well as a non-conservative radiation pressure torque on small particles with anisotropic electric susceptibility tensor. The non-conservative torque can angularly accelerate the nanoparticles, as has been observed in several experiments \cite{arita2013,kuhn2017,kuhn2017part2,reimann2018,ahn2018}  up to GHz rotation frequencies. Microscopically, the radiation pressure torque is caused by scattering of photons and thus always implies heating of the rotation state \cite{stickler2016,stickler2016part2}. Elliptic coherent scattering cooling balances this heating by cooling the rotations via two orthogonal cavity modes.

In this paper, we present the theoretical framework of elliptic coherent scattering cooling. Specifically, we provide three theoretical tools for describing future experiments with levitated nanorotors: First, we derive the conservative optical potential and the non-conservative radiation pressure torque exerted by an elliptically polarized laser on aspherical particles. Second, we present the master equation of the combined nanoparticle-cavity motion and discuss how quantum cooling is affected by the particle shape and laser ellipticity.Third, we derive the resulting power spectral densities of the cavity output modes, containing signatures of the rotational non-linearities if the particle is not deeply trapped. We expect the here presented theoretical framework will be instrumental for devising and interpreting opto-mechanical experiments with aspherical nanoparticles at the quantum limit.

\section{Radiation pressure force and torque}\label{sec:2}

\subsection{Electric field integral equation}\label{electricfieldsection}

This section derives the radiation pressure torque exerted by an elliptically polarised laser beam of wave number $k=\omega/c$ onto an aspherical nanoparticle. Specifically, we consider an  ellipsoidally shaped dielectric with mass $m$, relative permittivity $\varepsilon$, and principal diameters $\ell_a<\ell_b<\ell_c$ much smaller than the laser wavelength. It has the volume $V = \pi \ell_a \ell_b \ell_c/6$ and the moments of inertia $I_a = m(\ell_b^2+\ell_c^2)/20$, $I_b = m(\ell_a^2+\ell_c^2)/20$ and $I_c = m(\ell_a^2+\ell_b^2)/20$. 

To calculate the radiation pressure torque exerted by an incoming laser beam ${\bf E}({\bf r})e^{-i\omega t}$, we must solve Maxwell's equations in the presence of the dielectric. Since the nanoparticle is much smaller than the laser wavelength, the polarization field inside the particle can be approximated as being spatially homogeneous and rotated according to the particle susceptibility. For an ellipsoidal particle at center-of-mass position ${\bf R}=(x,y,z)$ and with orientation $\Omega$, the induced dipole moment is approximately given by $\mathbf{p}_{\rm RG}({\bf R},\Omega)\simeq \varepsilon_0V \chi(\Omega)\mathbf{E}(\mathbf{R})$ (Rayleigh-Gans approximation) \cite{hulst1981}, with the orientation-dependent susceptibility tensor $\chi(\Omega) = R(\Omega)\chi_0 R^{ T}(\Omega)$. Here, $R(\Omega)$ denotes the rotation matrix transforming between the body-fixed frame and the laboratory frame, see App.\,\ref{app:eulerangles}.

The susceptibility tensor is real and diagonal in the body fixed frame with eigenvalues $\chi_0 = \text{diag}(\chi_a,\chi_b,\chi_c)$ with $\chi_l = (\varepsilon - 1)/[1+(\varepsilon-1)N_l]$ being anisotropic because of the aspherical shape of the nanoparticle. The shape-induced optical anisotropy is quantified by the dimensionless depolarization factors
\begin{align}\label{depol}
	N_l = \frac{\ell_ a \ell_b \ell_c}{2}\int_{0}^{\infty} \frac{\text{d}s}{(s + \ell_l^2)\sqrt{(s + \ell_a^2)(s+\ell_b^2)(s+\ell_c^2)}},
\end{align}
implying $\chi_a < \chi_b < \chi_c$ and $\sum_l N_l = 1$.
	
The induced dipole moment in the Rayleigh-Gans approximation yields the conservative optical force and torque, see below. However, to determine the radiation pressure torque we need corrections in the first order of the particle volume $V k^3$. These corrections can be calculated from the electric field integral equation for the internal electric field ${\bf E}_{\rm int}$,
\begin{widetext}
\begin{align}\label{internalfield}
	\mathbf{E}_{\rm int}(\mathbf{r}) =& \mathbf{E}(\mathbf{r}) + \frac{\varepsilon - 1}{4\pi} (\nabla\otimes\nabla + k^2)\int_{V({\bf R},\Omega)} \text{d}^3 r' \, e^{ik|\mathbf{r}-\mathbf{r}'|} \frac{\mathbf{E}_{\rm int}(\mathbf{r}')}{|\mathbf{r}-\mathbf{r}'|}.
\end{align}
Here, the integration covers the particle volume at position ${\bf R}$ and with orientation $\Omega$. This equation follows from Maxwell's equations for a dielectric illuminated by the incoming field ${\bf E}({\bf r})$ \cite{schiffer1979}. Here, $\otimes$ denotes the dyadic (exterior) product. Once the internal field has been determined, the external field ${\bf E}_{\rm ext}({\bf r})$ can be determined by evaluating \eqref{internalfield} for positions outside the particle volume. The resulting scattered fields are given by ${\bf E}_{\rm s}({\bf r}) = {\bf E}_{\rm ext}({\bf r}) - {\bf E}({\bf r})$.

To calculate corrections in the linear order of the particle volume, we separate off the Rayleigh-Gans approximation by employing the ansatz  $\mathbf{E}_{\rm int} = \chi \mathbf{E}/(\varepsilon-1) + \mathbf{E}_{\rm cor}$ for the internal field. Treating these corrections ${\bf E}_{\rm cor}$ as small and expanding the Eq.\,(\ref{internalfield}) to the third order in $k$, yields the following equation
\begin{align}\label{eq:3}
	\mathbf{E}_{\rm cor}(\mathbf{r}) \simeq & \frac{iVk^3}{6\pi}\chi\mathbf{E}(\mathbf{r}) + \frac{\varepsilon - 1}{4\pi} \int_{V({\bf R},\Omega)} \text{d}^3 r'  \left [ 3 (\mathbf{r}-\mathbf{r}')\otimes (\mathbf{r}-\mathbf{r}') - \mathds{1}|\mathbf{r}-\mathbf{r}'|^2 \right ] \frac{\mathbf{E}_{\rm cor}(\mathbf{r}')}{|\mathbf{r}-\mathbf{r}'|^5}.
\end{align}
\end{widetext}
Here we used that all terms linear in $k$ vanish and that those quadratic in $k$ amount to small corrections of the conservative optical potential (see App.\,\ref{app:d}), which can be neglected. The third-order contributions, on the other hand, can give rise to non-conservative forces, as discussed below.

In lowest order of $V k^3$, the solution of Eq.\,\eqref{eq:3} is given by $\mathbf{E}_{\rm cor}(\mathbf{r}) \simeq i V k^3 \chi^2 \mathbf{E}(\mathbf{r})/6\pi(\varepsilon -1)$. Combining this correction with the Rayleigh-Gans approximation yields the induced electric dipole moment 
\begin{align}\label{dipolemoment}
	\mathbf{p}({\bf R},\Omega)\simeq \varepsilon_0 V \chi(\Omega)\left(1 + \frac{iVk^3}{6\pi}\chi(\Omega)\right)\mathbf{E}(\mathbf{R}).
\end{align}
Even though the correction is suppressed by $Vk^3\chi_c$, it can lead to a significant torque, as  shown next.
	
\subsection{Optical forces and torques}\label{opticalforcessection}

The total dipole force and torque acting on a dielectric particle can be expressed by integrating Maxwell's stress tensor $T$ over a spherical surface of infinite radius \cite{jackson,novotny2012principles},
\begin{subequations}\label{eq:Maxwellintegral}
\begin{align}
	\mathbf{F} = \lim\limits_{r'\rightarrow\infty} r'^2 \int \text{d}^2\mathbf{n}\, T(r' \mathbf{n})\mathbf{n},
\end{align}
and
\begin{align}
	\mathbf{N} = \lim\limits_{r'\rightarrow\infty} r'^3 \int \text{d}^2\mathbf{n}\, \mathbf{n}\times \left[T(r' \mathbf{n})\mathbf{n}\right].
\end{align}
\end{subequations}

The tensor $T$ is quadratic in the electric and the magnetic field, which in turn consist of the incident and the scattering fields. Since the contribution of the incident fields does not lead to a force or torque, Eqs. (\ref{eq:Maxwellintegral}) involve two contributions. First, the interference term between the incident and the scattered fields gives rise to the dipole force and torque \cite{novotny2012principles}
\begin{subequations}\label{forceandtorque}
	\begin{align}
		\mathbf{F}_{\rm dip} = \frac 1 2 \text{Re} \left [\mathbf{p}^* \cdot \left(  \nabla \otimes \mathbf{E} \right)^T \right ],
	\end{align}
and
\begin{align}
		\mathbf{N}_{\rm dip} = \frac 1 2 \text{Re} \left( \mathbf{p}^* \times \mathbf{E} \right).
	\end{align}
\end{subequations}
Second, the stress tensor of the pure scattering field describes the interaction between different volume elements within the particle leading to an additional torque $\mathbf{N}_{\rm s}$ (the corresponding force vanishes).

Asymptotically expanding the scattered fields $\mathbf{E}_{\rm s}$ and $\mathbf{B}_{\rm s}$ in $1/r'$ as $\mathbf{E}_{\rm s} = \mathbf{E}_{1/r'} + \mathbf{E}_{1/r'^2} + O(1/r'^3)$, and likewise for $\mathbf{B}_{\rm s}$, and using the transversality of electromagnetic radiation, $\mathbf{n} \cdot \mathbf{E}_{1/r'} = 0$ and $\mathbf{n}\cdot\mathbf{B}_{1/r'} = 0$, yields
\begin{align}\label{selftorque}
	\mathbf{N}_{\rm s} =& \lim\limits_{r'\rightarrow\infty}\frac{r'^3}{ 2 } \text{Re}\biggl[ \int \text{d}^2\mathbf{n}\, \varepsilon_0 \left( \mathbf{n}\cdot \mathbf{E}_{1/r'^2}^* \right)\mathbf{n}\times\mathbf{E}_{1/r'} \nonumber\\ &+ \frac{1}{\mu_0} \left( \mathbf{n}\cdot \mathbf{B}_{1/r'^2}^* \right)\mathbf{n}\times\mathbf{B}_{1/r'} \biggr]
\end{align}
Here, all fields are evaluated at $r'\mathbf{n}$ and a time average over one optical period is performed. The scattered fields in \eqref{selftorque} follow from the electric field integral equation (\ref{internalfield}) for the external field and Maxwell's equations as 
\begin{subequations}
	\begin{align}
		\mathbf{n}\cdot\mathbf{E}_{1/r'^2} &= -\frac{ik e^{ikr'}}{2\pi\varepsilon_0 r'^2} \mathbf{n}\cdot\mathbf{p}\, e^{-ik\mathbf{n}\cdot \mathbf{r}},
	\end{align}
and
	\begin{align}
		\mathbf{n}\times\mathbf{E}_{1/r'} &= -\frac{k^2 e^{ikr'}}{4\pi\varepsilon_0 r'} \mathbf{p}\times\mathbf{n} \, e^{-ik\mathbf{n}\cdot\mathbf{r}},
	\end{align}
\end{subequations}
together with $\mathbf{n}\cdot \mathbf{B}_{1/r'^2} = 0$.

\begin{figure*}[tp]
    \centering
    \includegraphics[width = 0.98\textwidth]{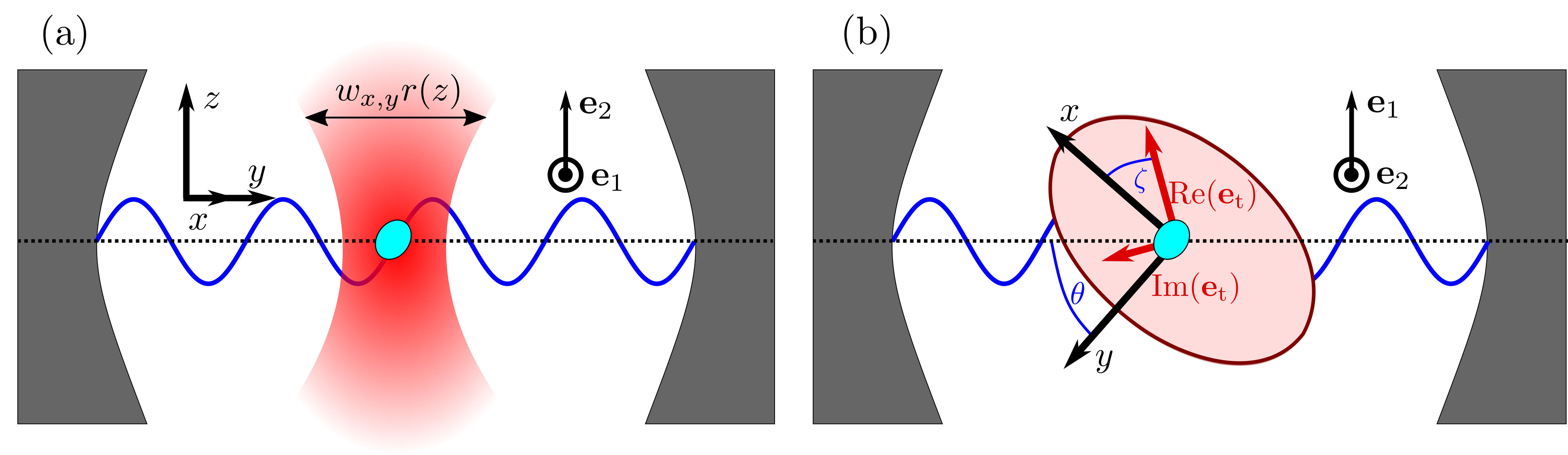}
    \caption{(a) Side view and (b) top view of the setup. An aspherical particle is trapped by an elliptically polarised and shaped tweezer between two cavity mirrors. Coherent scattering of tweezer photons into two orthogonal cavity modes couples the rotational and translational particle motion to the dissipative cavity dynamics. The tweezer propagates in $z$-direction, and the main tweezer axes are $x$ and $y$ with $w_{x} r(z)$ and $w_y r(z)$ being the major and minor axes of the elliptic cross section, as given by the broadening function $r(z)$. The elliptic tweezer polarization,  denoted by ${\bf e}_t$, is rotated by the angle $\zeta$ with respect to the main tweezer axes, whereas the cavity axis is rotated by $\theta$. The linear polarization vectors of the two orthogonal cavity modes are denoted by ${\bf e}_1$ and ${\bf e}_2$.}
    \label{fig:sketch}
\end{figure*}

Inserting these expressions into (\ref{selftorque}) and adding them to the dipole force and torque (\ref{forceandtorque}) yields the total force and torque acting on the particle
\begin{subequations}\label{fnt}
	\begin{align}
		\mathbf{F} = &\nabla\left( \frac{\varepsilon_0 V}{4} \mathbf{E}^* \cdot \chi \mathbf{E} \right)\\&+ \frac{\varepsilon_0 k^3 V^2}{12\pi}\text{Im} \left[ (\chi\mathbf{E}^*) \cdot [ \nabla \otimes (\chi\mathbf{E})]^T \right],\nonumber
	\end{align}
and
	\begin{align}
		\mathbf{N} =& \frac{\varepsilon_0 V}{2} \text{Re} \left[ (\chi\mathbf{E}^*)\times\mathbf{E} \right]\\ &+ \frac{\varepsilon_0 k^3 V^2}{12\pi}\text{Im} \left[ (\chi^2\mathbf{E}^*)\times\mathbf{E} - (\chi\mathbf{E}^*)\times(\chi\mathbf{E}) \right].\nonumber
	\end{align}
\end{subequations}
Here, the field is  evaluated at the particle position $\mathbf{R}$. Terms of order higher than  $Vk^3$ are neglected.

Equations \eqref{fnt} describe the conservative and non-conservative forces and torques acting on small but ellipsoidally shaped dielectrics in a light field of arbitrary polarization. The first terms can be associated with a single conservative optical potential (see below). Note that the net torque vanishes for an isotropic particle, which would not be the case if the self-interaction contribution to the scattering torque had been neglected. 

The radiation pressure torque for circularly polarised tweezers has been experimentally observed to  accelerate aspherical nanoparticles angularly, up to GHz rotation rates \cite{kuhn2017,kuhn2017part2,ahn2018,reimann2018}. Its interplay with the radiation pressure force leads to strong correlations between the center-of-mass and rotational dynamics \cite{arita2020}, with great potential for precision sensing. In the quantum regime, the non-conservative radiation pressure force and torque are accompanied by motional heating of the particle \cite{papendell2017}.

\section{Coupled cavity-particle dynamics}\label{dynamics}

\subsection{Tweezer and cavity fields}

In the setup of coherent scattering, the nanoparticle is trapped by a tweezer with optical frequency $\omega$ (wavenumber $k = \omega/c$) inside a cavity with two orthogonal cavity modes of frequency $\omega_{\rm c}$, assumed degenerate for simplicity.
The incoming electric field $\mathbf{E}(\mathbf{r})e^{-i\omega t}$ seen by the particle is  the sum of all fields acting on the dielectric,
\begin{align} \label{eq:field}
	\mathbf{E}(\mathbf{r}) = \sqrt{\frac{2\hbar\omega}{\varepsilon_0 V_{\rm c}}} \left [\epsilon \mathbf{e}_{\rm t} f_{\rm t}(\mathbf{r})  + \sum_{j=1,2} b_j \mathbf{e}_j f_{\rm c}(\mathbf{r}) \right ].
\end{align}
The cavity mode volume $V_{\rm c} = \pi L_{\rm c} w_{\rm c}^2/4$ is determined by the cavity length $L_{\rm c}$ and the cavity beam waist $w_{\rm c}$. The cavity and tweezer mode amplitudes are described by the complex dimensionless mode variables $b_{1,2}$ and $\epsilon$, respectively. The latter is chosen to be real and related to the tweezer power $P_{\rm t}$ and tweezer waists $w_{x,y}$ as \cite{gonzalez2019}
\begin{align}
	\epsilon = \sqrt{\frac{2P_{\rm t}k V_{\rm c}}{\pi \hbar\omega^2 w_x w_y}}.
\end{align}

The tweezer mode function is well approximated by a traversing Gaussian beam with propagation direction ${\bf e}_z$ and intensity main axes as ${\bf e}_{x,y}$ \cite{gonzalez2019}
\begin{subequations}
	\begin{align}
		f_{\rm t}(\mathbf{r}) = \frac{1}{r(z)} \exp \left (-\frac{x^2}{w_x^2 r^2(z)}- \frac{y^2}{w_y^2 r^2(z)}\right ) e^{i[kz - \phi_{\rm t}(\mathbf{r})]},
	\end{align}
with the tweezer Rayleigh range $z_{\rm R} \simeq k w_x w_y/2$, the dimensionless broadening function $r(z)\simeq\sqrt{1+z^2/z_{\rm R}^2}$, and the tweezer Gouy-phase
\begin{align}
	\phi_{\rm t}(\mathbf{r}) \simeq \arctan\left(\frac{z}{z_{\rm R}}\right) - \frac{kz}{2} \frac{x^2+y^2}{z^2+z_{\rm R}^2}.
\end{align}
The tweezer is elliptically polarized, $\mathbf{e}_{\rm t} = \cos\psi\, \mathbf{e}_{\rm t,1} + i\sin\psi\, \mathbf{e}_{\rm t,2}$, as described by the ellipticity $\psi\in [0,\pi/4]$. For $\psi=0$ the tweezer polarization is linear, and for $\psi=\pi/4$ it is circular. The polarization axes $\mathbf{e}_{\rm t,1}=\cos\zeta \mathbf{e}_x -\sin\zeta \mathbf{e}_y$ and $\mathbf{e}_{\rm t,2}=\sin\zeta \mathbf{e}_x +\cos\zeta \mathbf{e}_y$ are described by the rotation angle $\zeta$ relative to the intensity main axes, see Fig.\,\ref{fig:sketch}.

Aligning the cavity axis orthogonal to the tweezer propagation direction allows one to express the two orthogonal cavity mode polarizations as $\mathbf{e}_1 = \cos\theta {\bf e}_x -\sin\theta {\bf e}_y$ and $\mathbf{e}_2 = \mathbf{e}_z$ (see Fig.~\ref{fig:sketch}).
The cavity mode functions are well approximated by standing wave Gaussian beams,
	\begin{align}
		f_{\rm c}(\mathbf{r})= \cos\left[k\left( \mathbf{e}_2 \times \mathbf{e}_1 \right)\cdot \mathbf{r} + \phi \right]\exp \left (-\frac{(\mathbf{e}_1 \cdot \mathbf{r})^2 + z^2}{w_{\rm c}^2} \right ).
	\end{align}
\end{subequations}
Here, we neglected the Gouy-phase and the broadening factor for the cavity modes since their Rayleigh range is typically several orders of magnitude larger than all other length scales.

\subsection{Cavity equations of motion}

The coupled dynamics of the nanoparticle and the cavity field can be obtained by combining the force and torque \eqref{fnt} with the equations of motion for the cavity modes $b_{1,2}$  \cite{salzburger2009}. For a particles small compared to the optical wave length, the equations of motion for the cavity modes $b_{1,2}$ follow as
\begin{align}\label{beq}
	\dot b_j = (i\Delta-\kappa)b_j + \frac{i\omega}{2\sqrt{2\hbar\omega\varepsilon_0 V_{\rm c}}} \mathbf{p}\cdot \mathbf{e}_j f_{\rm c} (\mathbf{R}),
\end{align}
where $\Delta = \omega-\omega_{\rm c}$ is the tweezer detuning  and $\kappa$ the cavity loss rate due to the imperfect reflectivity of the cavity mirrors.
	
Using the induced dipole moment (\ref{dipolemoment}) and defining the complex vector $b=(b_1,b_2)$, the cavity equations of motion take the form
\begin{align}\label{eq:cavity}
	\dot b = A(\mathbf{R},\Omega)b + \eta(\mathbf{R},\Omega),
\end{align}
with the $\mathbb{C}^2$-matrix $A(\mathbf{R},\Omega) = i\Delta_{\rm eff}(\mathbf{R},\Omega) - \kappa_{\rm eff}(\mathbf{R},\Omega)$. It contains the effective detuning matrix
\begin{subequations}\label{eq:cavpara}
\begin{align}
	[\Delta_{\rm eff}(\mathbf{R},\Omega)]_{jj'} = \Delta \delta_{jj'} - U_0  \mathbf{e}_j \cdot \chi(\Omega) \mathbf{e}_{j'} f_{\rm c}^2(\mathbf{R}),
\end{align}
and the effective damping matrix
\begin{align}
	[\kappa_{\rm eff}(\mathbf{R},\Omega)]_{jj'} = \kappa\delta_{jj'} + \frac{\gamma_{\rm sc}}{2} \mathbf{e}_j \cdot \chi^2(\Omega) \mathbf{e}_{j'} f_{\rm c}^2(\mathbf{R}),
\end{align}
with the coupling frequency $U_0 = -\omega V/2V_{\rm c}$ and the scattering rate $\gamma_{\rm sc} = \omega k^3 V^2/6\pi V_{\rm c}$. The matrix $\Delta_{\rm eff}$ is real and symmetric and $\kappa_{\rm eff}$ is real, symmetric, and positive definite, implying that $A^{-1}$ is well-defined.

The cavity modes are driven by the scattering of tweezer light off the nanoparticle, as described by the pump vector
\begin{align} \label{eq:drive}
	[\eta(\mathbf{R},\Omega)]_j = -\epsilon  \mathbf{e}_j \cdot\left(iU_0 \chi(\Omega) + \frac{\gamma_{\rm sc}}{2} \chi^2(\Omega)\right)\mathbf{e}_{\rm t}  f_{\rm c}(\mathbf{R}) f_{\rm t}(\mathbf{R}).
\end{align}
\end{subequations}
The induced dipole moment in the dielectric thus effectively shifts the cavity detuning, adds dissipation due to scattering into free space, and drives the cavity mode. While genuine scattering $\gamma_{\rm sc}$  barely contributes to the drive \eqref{eq:drive} for realistic dielectrics, the cavity is mainly pumped by the coherently oscillating polarization density in the particle, referred to as {\it coherent scattering} \cite{delic2019part2,windey2019}.

\subsection{Nanoparticle-cavity Hamiltonian}

We can identify the total  Hamiltonian of the coupled motion of the nanoparticle and the cavity by putting aside the damping and scattering contributions to the cavity dynamics,
\begin{align}\label{hamiltonian}
	H = H_0 + V_{\rm opt} - \hbar\Delta\left(b_1^* b_1 + b_2^* b_2 \right).
\end{align}
The free nanoparticle Hamiltonian $H_0$ reads, in terms of the center-of-mass position ${\bf R}$, the Euler angles $\Omega = (\alpha,\beta,\gamma)$ in the $z$-$y'$-$z''$ convention, and the corresponding canonical momenta $p_q\in(\mathbf{p},p_{\Omega})$, see App.\,\ref{app:eulerangles}, as follows:
\begin{align}\label{eq:h0}
 H_0 = & \frac{1}{2 I_a}\left[ \cos\gamma\frac{p_\alpha-p_\gamma\cos\beta}{\sin\beta} - \sin\gamma p_\beta \right]^2\nonumber\\ 
 &+ \frac{1}{2 I_b}\left[\sin\gamma \frac{p_\alpha-p_\gamma\cos\beta}{\sin\beta} +\cos\gamma p_\beta \right]^2\nonumber\\ &+  \frac{p_\gamma^2}{2 I_c} + \frac{\mathbf{p}^2}{2m}.
\end{align}
The conservative optical potential
\begin{equation}\label{eq:optpot}
	V_{\rm opt} (\mathbf{R},\Omega) = - \frac{\varepsilon_0 V}{4} \mathbf{E}^* (\mathbf{R})\cdot \chi(\Omega)\mathbf{E} (\mathbf{R}).
\end{equation}
describes the conservative part of the optical force and torque (\ref{fnt}) according to $\dot p_q = -\partial_q H + F_q^{\rm rad}$, where $q\in(\mathbf{R},\Omega)$. The non-conservative radiation pressure force and torques are as follows:
\begin{align}\label{radforce}
	F_q^{\rm rad} (\mathbf{R},\Omega) = \frac{\varepsilon_0 V^2 k^3}{12\pi}\text{Im} \left [\mathbf{E}^*({\bf R})\cdot \chi(\Omega) \partial_q [\chi(\Omega)\mathbf{E}({\bf R})] \right].
\end{align}
The Hamiltonian (\ref{hamiltonian}) and Eqs. (\ref{radforce}) and \eqref{eq:cavpara} fully describe the coupled nanoparticle-cavity dynamics in the elliptic coherent scattering setup.
	
\section{Quasi-static cavity cooling}

In the \emph{bad cavity regime}, when the cavity decay is much faster than the mechanical  timescale, $1/\kappa \ll \tau_{\rm m}$, 
the field reacts almost instantaneously to the nanoparticle dynamics. Likewise, if the cavity dynamics is much faster than the mechanical motion, $1/|\Delta| \ll \tau_{\rm m}$, the cavity field follows the nanoparticle motion adiabatically. 

In both cases, the resulting quasi-static field dynamics can be determined by writing $b = b_{\rm s} + \Delta b $, where $b_{\rm s} = -A^{-1} \eta$ is the stationary solution of the cavity equations of motion at fixed  nanoparticle position and orientation.
If the particle moves the deviation $\Delta b$ evolves as
\begin{align}
	\frac{\rm d}{{\rm d} t}\Delta{b}=
	%A\Delta b-\frac{\rm d}{{\rm d} t}{b}_{\rm s}=
	A\Delta b\ -\sum_{q}{\dot{q}\,\partial_q b_{\rm s}},
\end{align}
depending on the particle  velocities $\dot{q}$. Assuming that the field reacts instantaneously on the particle dynamics, we can approximate $\Delta\dot{b}\simeq 0$, which yields the quasi static field
\begin{align}\label{eq:bqs}
	 b\simeq b_{\rm s} + A^{-1}\sum_{q}\dot{q}\,\partial_qb_{\rm s}.
\end{align}
Its dependence on the particle velocity can give rise to cavity cooling, as we will show next.

The  quasi-static cavity field exerts a friction force $F_q^{\rm fric}$, which is linear in the nanoparticle velocities. This  follows from inserting the total field \eqref{eq:field} with the cavity amplitudes  (\ref{eq:bqs}) into the total optical force and torque $F_q = - \partial_q V_{\rm opt} + F_q^{\rm rad}$. Keeping only terms linear in $\Delta b=b-b_{\rm s}$ yields
\begin{align}
	F_q^{\rm fric}\simeq& -2\hbar\, \text{Im}\sum_{j,j' = 1,2} \Delta b_j^{*} A_{jj'} \partial_q b_{{\rm s},j}  \\ &+ 2\hbar\gamma_{\rm sc}  \,\text{Im}\,\sum_{j = 1,2} \Delta b_j^* \epsilon f_{\rm c}\chi \mathbf{e}_j\cdot \partial_q (f_{\rm t} \chi \mathbf{e}_{\rm t}) \nonumber\\ &+  2\hbar\gamma_{\rm sc} \,\text{Im}\,\sum_{j,j' = 1,2} \Delta b_j^* b_{{\rm s},j} f_{\rm c}\chi \mathbf{e}_{j}\cdot \partial_q (f_{\rm c} \chi \mathbf{e}_{j'}).\nonumber
\end{align}
For  particles small compared to the optical wave length the radiation pressure contribution to this friction force can be neglected because it is suppressed by $V k^3$.

We proceed to calculate the phase-space contraction rate in order to quantify the extent to which the particle state of motion is damped. As shown in App.\,\ref{phasespace}, the total phase-space contraction rate is $\Gamma_{\rm c} = -\sum_{q} \partial_{p_q} F^{\rm fric}_q$. By neglecting the radiation pressure and inserting $\dot q = \partial H_0/\partial p_q$ one obtains
\begin{align}\label{contraction}
	\Gamma_{\rm c} \simeq 2\hbar\,{\rm Im}\sum_{qq'jj'} \left(\partial_{q}b_{{\rm s},j} ^*\right)\frac{\partial^2H_0}{\partial p_q \partial p_{q'}} B_{jj'} \, \partial_{q'} b_{{\rm s},j'} 
\end{align}
with $B=\left(A^{-1}\right)^\dagger A$. This is a sesquilinear form in $\partial_{q} b_{{\rm s},j}$ in terms of the real and symmetric matrix $  [\partial^2 H_0/\partial p_q \partial p_{q'}]_{qq'}  \text{Im}(B)_{jj'}$. Since the Hessian of $H_0$ is positive, the rate $\Gamma_{\rm c}$ is strictly positive (negative) if $\text{Im}(B)$ is positive (negative).

The eigenvalues of  $A$ are given by  $i\lambda-\kappa$, with
\begin{align}
\lambda =&  \Delta-U_0 f_{\rm c}^2 \biggl[ \frac{\mathbf{e}_1\cdot \chi \mathbf{e}_1 + \mathbf{e}_2\cdot \chi \mathbf{e}_2}{2}\nonumber\\ &\pm \sqrt{\frac{(\mathbf{e}_1\cdot \chi \mathbf{e}_1 - \mathbf{e}_2\cdot \chi \mathbf{e}_2)^2}{4} + (\mathbf{e}_1\cdot \chi \mathbf{e}_2)^2} \biggr],
\end{align}
so that $\text{Im}(B)$ has the eigenvalues $-2\kappa\lambda/(\kappa^2+\lambda^2)$. The contraction rate $\Gamma_{\rm c}$ is therefore  positive for all $q$ provided all eigenvalues $\lambda$ of $\Delta_{\rm eff}$ are negative for all $q$. This is guaranteed by $\Delta< U_0 ( \chi_c + \chi_b)$, i.e.\ if the tweezer is sufficiently far red detuned. It then follows that the classical ro-translational motion of the particle cools down to ever lower temperatures, as long as the radiation pressure can be neglected. 

The quasi-static phase space contraction rate (\ref{contraction}) quantifies cooling of the full nanoparticle rotational and translational motion due to the retarded back-action of the cavity. It is valid in both the adiabatic-cavity regime and the bad-cavity regime. Since it  takes the full non-linearity of the electric field and of the rotation dynamics into account, the rate applies even if the particle is not deeply trapped. It is thus also relevant for experiments aiming at transit cooling \cite{asenbaum2013} or dispersive capturing \cite{stickler2016} of nanoparticles.
	
\section{Quantum dynamics}\label{quantumdynamics}

\subsection{Markovian master equation}

The quantum dynamics of the combined nanoparticle-cavity state may be described by a completely positive Markovian quantum master equation for the density operator $\rho$. Promoting the phase-space coordinates of the rigid rotor and the mode variables of the cavity field to operators (subject to the canonical commutation relations and $[b,b^\dagger]=1$, see App.~\ref{app:eulerangles}), and adding the Lindblad dissipators for cavity loss, Rayleigh scattering of photons \cite{papendell2017}, and gas scattering, one arrives at
\begin{align}\label{mastereq}
	\partial_t \rho =& -\frac{i}{\hbar} \left[ H, \rho \right] + 2\kappa \sum_{j=1,2} \left( b_j\rho b_j^\dagger - \frac 1 2 \lbrace b_j^\dagger b_j,\rho\rbrace\right)\nonumber\\
	&+ \mathcal{L}_{\rm gas}\rho  + \int_{S_2} \text{d}^2\mathbf{n} \sum_s \left[ L_{\mathbf{n}s}\rho L_{\mathbf{n}s}^\dagger - \frac 1 2 \lbrace  L_{\mathbf{n}s}^\dagger L_{\mathbf{n}s}, \rho\rbrace \right].
\end{align}
Here, $H$ is the quantized version of \eqref{hamiltonian}, including the quantum potential \eqref{eq:quantpot}. The Lindblad superoperator $\mathcal{L}_{\rm gas}$ describing collisional decoherence due to gas scattering is given in Ref. \cite{stickler2016part2}; it leads to linear and angular momentum diffusion of the particle \cite{papendell2017}.

The last term on the right-hand side of Eq.~\eqref{mastereq} describes the non-conservative optical force and torque, and the resulting decoherence due to Rayleigh scattering. The integral over the unit sphere accounts for scattering into all possible directions $\mathbf{n}$ and the sum covers the possible photon polarizations $\mathbf{t}_s$. The associated Lindblad operators $L_{\mathbf{n}s}$ can be determined with the {\em monitoring approach} in the Rayleigh-Gans approximation \cite{stickler2016part2}, 
\begin{align}\label{lindbladian}
	L_{\mathbf{n}s} = \sqrt{\frac{\varepsilon_0 k^3}{2\hbar}}\frac{V}{4\pi} \mathbf{t}_s^*\cdot \chi(\Omega)\mathbf{E}(\mathbf{R})   e^{-ik\mathbf{n}\cdot\mathbf{R}}.
\end{align}

They are diagonal in the rotor orientation $\Omega$ and the position $\mathbf{R}$.
The total power scattered from the nanoparticle into free space,  calculated as $P_{\rm sc} = \hbar \omega \int \text{d}^2\mathbf{n} \sum_s \langle L_{\mathbf{n}s}^\dagger L_{\mathbf{n}s} \rangle$, is consistent with the result obtained from  Poynting's theorem with the scattered fields given in Sec. \ref{sec:2}. The Ehrenfest equations obtained from \eqref{mastereq} for the operators $q$ and $b_{1,2}$ yield the classical equations of motion, Sec.\,\ref{dynamics}, including the radiation pressure (\ref{radforce}), when replacing all operators by their expectation values. In addition, the master equation (\ref{mastereq}) describes decoherence and diffusion of the quantum state.

\subsection{Deep-trapping regime}

We call the nanoparticle \emph{deeply trapped} if its 6D motion remains sufficiently close to a stable equilibrium of the optical trap, so that all forces and torques can be linearized.
For elliptical tweezer polarization the  minimum $q_{\rm tw} = ({\bf R}_{\rm tw}, \Omega_{\rm tw})$ of the bare tweezer potential is located at $\mathbf{R}_{\rm tw}= (0,0,0)$ and $\Omega_{\rm tw} = (-\zeta,\pi/2,0)$. An elliptically polarised tweezer provides an orientational minimum in all angular degrees of freedom because the two main axes of the particle tend to align with the two different polarization axes, which then fixes the third axis. The non-conservative force and torque can  shift this minimum slightly, but this effect is tiny for small particles, since it is proportional to $V k^3 \chi_c$. For the particles considered below and in Ref.~\cite{prlsubmission}, this amounts to at most a few nanometers and arcseconds.

Also the cavity field will in general exert a force and torque since the mode occupations are nonzero at the tweezer minimum $b_{\rm tw} = b_{\rm s}(q_{\rm tw})$. (They would vanish only in the limit of  far detuning.)
To quantify the effect of the   cavity-induced optical forces,  we harmonically expand the Hamiltonian along the flat configuration space tangent to the tweezer minimum, assign the metric determinant to the quantum state, and neglect the quantum potential, see App.~\ref{app:eulerangles}.
The  Hilbert space is then based on unbounded translation and libration coordinates %$({\bf r},\Omega)\in \mathbb{R}^6$  
with a flat metric,  ${\rm d}^3{\bf R} {\rm d}^3\Omega = {\rm d}x {\rm d}y {\rm d}z {\rm d}\alpha {\rm d}\beta {\rm d}\gamma$. The  master equation for the deep trapping regime can then be obtained by harmonically expanding in all remaining coordinate operators around the tweezer minimum.

Specifically, to derive the harmonically expanded Hamiltonian, we first approximate the tweezer mode function as
\begin{subequations}\label{eq:harm}
\begin{align}
f_{\rm t}({\bf R}) \simeq & 1-\frac{x^2}{w_x^2}-\frac{y^2}{w_y^2} - \frac{z^2}{2z_{\rm R}^2}\nonumber\\ &+ i\left(k-\frac{1}{z_{\rm R}}\right)z -\frac 1 2 \left( k-\frac{1}{z_{\rm R}} \right)^2 z^2,
\end{align}
and the cavity mode function as
\begin{align}
f_{\rm c}({\bf R}) \simeq \cos\phi &-  \sin\phi k (\mathbf{e}_2 \times \mathbf{e}_1)\cdot \mathbf{R} \nonumber\\ &- \frac 1 2 \cos\phi k^2\left[ (\mathbf{e}_2 \times \mathbf{e}_1)\cdot \mathbf{R} \right]^2 .
\end{align}
In a similar fashion, the susceptibility tensor can be expanded as
\begin{align}\label{suscapprox}
\chi(\Omega) \simeq & \chi(\Omega_{\rm tw}) + (\chi_c - \chi_b)  (\mathbf{e}_{\rm t,1} \otimes \mathbf{e}_{\rm t,2} + \mathbf{e}_{\rm t,2} \otimes \mathbf{e}_{\rm t,1}) \delta\alpha \nonumber \\ 
&- (\chi_c - \chi_a)  (\mathbf{e}_{\rm t,1} \otimes \mathbf{e}_z + \mathbf{e}_z \otimes \mathbf{e}_{\rm t,1})\delta\beta \nonumber\\ &+ (\chi_b - \chi_a)  (\mathbf{e}_{\rm t,2} \otimes \mathbf{e}_z + \mathbf{e}_z \otimes \mathbf{e}_{\rm t,2})\gamma \nonumber\\
&+(\chi_c-\chi_b)(\mathbf{e}_{\rm t,2} \otimes \mathbf{e}_{\rm t,2} - \mathbf{e}_{\rm t,1} \otimes \mathbf{e}_{\rm t,1})\delta\alpha^2 \nonumber\\
&+(\chi_c-\chi_a)(\mathbf{e}_z \otimes \mathbf{e}_z - \mathbf{e}_{\rm t,1} \otimes \mathbf{e}_{\rm t,1})\delta\beta^2 \nonumber\\
&+(\chi_b-\chi_a)(\mathbf{e}_z \otimes \mathbf{e}_z - \mathbf{e}_{\rm t,2} \otimes \mathbf{e}_{\rm t,2})\gamma^2 \nonumber\\
&-(\chi_c-\chi_a)(\mathbf{e}_{\rm t,2} \otimes \mathbf{e}_z + \mathbf{e}_z \otimes \mathbf{e}_{\rm t,2})\delta\alpha\delta\beta \nonumber\\
&-(\chi_b-\chi_a)(\mathbf{e}_{\rm t,1} \otimes \mathbf{e}_z + \mathbf{e}_z \otimes \mathbf{e}_{\rm t,1})\delta\alpha\gamma \nonumber\\
&+(\chi_b-\chi_a)(\mathbf{e}_{\rm t,1} \otimes \mathbf{e}_{\rm t,2} + \mathbf{e}_{\rm t,2} \otimes \mathbf{e}_{\rm t,1})\delta\beta\gamma
\end{align}
\end{subequations}
with $\delta\alpha = \alpha - \alpha_{\rm tw}$ and $\delta \beta = \beta - \beta_{\rm tw}$.

Inserting these expansions into the Hamiltonian $H$ and keeping only terms up to quadratic order in $\delta q = q - q_{\rm tw}$ and $\delta b  = b - b_{\rm tw}$, yield an expression with terms both quadratic and linear  in all coordinates $q\in\{x,y,z,\alpha,\beta,\gamma\}$. The linear terms describe that the equilibrium configuration $(q_{\rm eq},b_{\rm eq})$ slightly deviates from the tweezer minimum $(q_{\rm tw},b_{\rm tw})$ due to the finite cavity populations. We therefore define the mechanical mode operators  with respect to the equilibrium configuration as
\begin{align}\label{linearizedhamiltonians}
    a_q = \frac{1}{2 q_{\rm zp}}\left(q - q_{\rm eq} + i \frac{p_q}{m_q \omega_q}\right) \,,
\end{align}
with  $q^2_{\rm zp} = {\hbar/2m_q \omega_q}$ and $m_q^{-1} = \partial_{p_q}^2 H_0(q_{\rm tw})$. Hence, $m_{x,y,z}=m$, $m_\alpha = I_a$, $m_\beta = I_b$ and $m_\gamma = I_c$.

The resulting  Hamiltonian decomposes into the sum $H = H_1 + H_2$ of operators $H_j$, each of which acts only on a subset of the total configuration space,
\begin{align}\label{eq:h1}
\frac{H_j}{\hbar} = &\sum_{q \in S_j} \omega_q a_q^\dagger a_q -\sum_{q \neq q' \in S_j} g_{qq'} (a_q + a_q^\dagger)(a_{q'} + a_{q'}^\dagger) \nonumber\\
& -\sum_{ q \in S_j}\left[g_{jq} \delta b_j \left(a_q + a_q^\dagger\right) + \text{h.c.}\right]-\Delta_j \delta b_j^\dagger \delta b_j.
\end{align}
Specifically, the Hamiltonian $H_1$ involves only  the cavity mode $\delta b_1$ and the mechanical degrees of freedom $S_1 = \{x,y,z,\alpha\}$, while  $H_2$ acts only on the cavity mode $\delta b_2$ and the mechanical degrees of freedom $S_2 = \{\beta, \gamma\}$. The equilibrium orientations of $\beta$ and $\gamma$ coincide with the tweezer minimum, while the cavity mode $b_2$ is empty in the steady state, as follows from a direct calculation. 
 
The trapping frequencies follow from the harmonic expansions \eqref{eq:harm} and can be  written compactly as $\omega_q^2 = \partial^2_q V_{\rm opt}(q_{\rm tw},b_{\rm tw})/m_q$. Similarly, the direct mechanical couplings due to the optical potential are $g_{qq'} = -q_{\rm zp} q'_{\rm zp} \partial_q \partial_{q'} V_{\rm opt}(q_{\rm tw}, b_{\rm tw})/\hbar$ and the light-mechanical coupling constants can be written as $g_{jq} = - q_{\rm zp}\partial_{b_j} \partial_q V_{\rm opt}(q_{\rm tw},b_{\rm tw})/\hbar$. The effective detunings of the cavity modes are given by $\Delta_j = [\Delta_{\rm eff}(q_{\rm tw})]_{jj}$. Explicit expressions for all coupling constants and trapping frequencies can be found in Appendix \ref{parameterlist}.

\subsection{Diffusive recoil and gas heating}

In the deep trapping regime, the master equation \eqref{mastereq} can be approximated by replacing the exact Hamiltonian \eqref{hamiltonian} by Eqs. \eqref{eq:h1} and by Taylor-expanding the dissipators in \eqref{mastereq} around the tweezer minimum. In the following, we perform this calculation for recoil heating with Lindblad operators \eqref{lindbladian}.

As the first step we use that tweezer shot noise dominates over cavity shot noise because the cavity modes are only weakly occupied, $\epsilon \gg |b_{\rm eq}|$. Neglecting the cavity contribution in Eq.~(\ref{lindbladian}), we use the expansion (\ref{suscapprox}) of the susceptibility tensor as well as
\begin{align}\label{expansionheat}
	f_{\rm t}(\mathbf{R}) e^{-ik\mathbf{n}\cdot\mathbf{R}} &\simeq 1 + i\left[\left(k-\frac{1}{z_{\rm R}}\right)\mathbf{e}_z - k\mathbf{n}\right]\cdot\mathbf{R}\nonumber\\ &- \frac 1 2 \left\lbrace\left[\left(k-\frac{1}{z_{\rm R}}\right)\mathbf{e}_z - k\mathbf{n}\right]\cdot\mathbf{R}\right\rbrace^2.
\end{align}
Here we neglected all terms from the tweezer envelope that cancel in the master equation.

Inserting these expansions into the master equation yields non-conservative contributions quadratic in all $\delta q$, which describe decoherence and diffusion, and in addition coherent radiation pressure forces in $z$ and $\alpha$, which are linear in $\delta q$. Carrying out the sum over polarizations and the integral over scattering directions, shows that the linear contributions can be neglected, as already discussed above.

The resulting recoil heating rates  for the translational degrees of freedom $q\in\{x,y,z\}$ take the form
\begin{subequations}\label{recoils}
\begin{align}
	\xi_q^{\rm rec} = & \frac{\gamma_{\rm sc} \epsilon^2}{5}k^2 q_{\rm zp}^2\left[ (\chi_c^2 \cos^2\psi + \chi_b^2\sin^2\psi)\left (2+ u \,\delta_{zq} \right ) \right. \nonumber \\
	& \left. - \chi_c^2 \cos^2\psi \,\delta_{xq}- \chi_b^2\sin^2\psi \,\delta_{yq} \right],
\end{align}
with $u = 5(1-1/k z_{\rm R})^2$. The rotational recoil rates  read as
\begin{align}
	\xi_q^{\rm rec} = \gamma_{\rm sc} \epsilon^2 q_{\rm zp}^2\Delta\chi_q^2 \left[1 - \sin^2\psi \, \delta_{\beta q} - \cos^2\psi \, \delta_{\gamma q} \right]
\end{align}
\end{subequations}	
for $q\in \{ \alpha,\beta,\gamma \}$. They involve the susceptibility anisotropy $\Delta\chi_\alpha = |\chi_b - \chi_c|$ and $\Delta\chi_\beta$ and $\Delta\chi_\gamma$ given by cyclic permutation. We note that the translational heating rates reduce to the known expressions \cite{gonzalez2019} for spherical particles.

The calculation for collisional decoherence follows the same lines \cite{romero2011,papendell2017}. The resulting heating rates can be calculated from the friction rates $\gamma_q^{\rm gas}$ and the gas temperature $T_{\rm g}$ as $\xi_q^{\rm gas} =  k_{\rm B} \gamma_q^{\rm gas} T_{\rm g}/\hbar \omega_q$. The gas friction rates in general depend on the particle shape \cite{martinetz2018}. As an approximation for ellipsoidal bodies, we use the formula for a sphere whose diameter is equivalent to the middle axis $\ell_b$ for all degrees of freedom \cite{martinetz2018}
\begin{align}
\gamma_q^{\rm gas} \simeq \frac{5p_{\rm g}\ell_b^2}{6m} \sqrt{\frac{2\pi\mu}{k_B T_{\rm g}}}.
\end{align}
Here, $p_{\rm g}$ is the gas pressure and $\mu$ is the mass of the impinging gas atom (assumed to be helium below).

Finally, combining this with the Hamiltonians \eqref{eq:h1} and neglecting all off-resonant terms, the  master equation for the deep-trapping regime can be obtained as follows:
\begin{align}\label{linmasterequation}
	&\partial_t \rho \simeq -\frac{i}{\hbar}\left[ H, \rho \right] + 2\kappa \sum_{j=1,2}\left( \delta b_j \rho \delta b_j^\dagger - \frac 1 2 \lbrace \delta b_j^\dagger \delta b_j,\rho \rbrace \right)\nonumber\\ &+ \sum_q \xi_q\biggl[ a_q \rho a_q^\dagger - \frac 1 2 \lbrace a_q^\dagger a_q,\rho \rbrace + a_q^\dagger \rho a_q - \frac 1 2 \lbrace a_q a_q^\dagger,\rho\rbrace \biggr],
\end{align}
with the total heating rates $\xi_q = \xi_q^{\rm rec} + \xi_q^{\rm gas}$.

\begin{figure*}[tp]
	\centering
	\includegraphics[width=.95\linewidth]{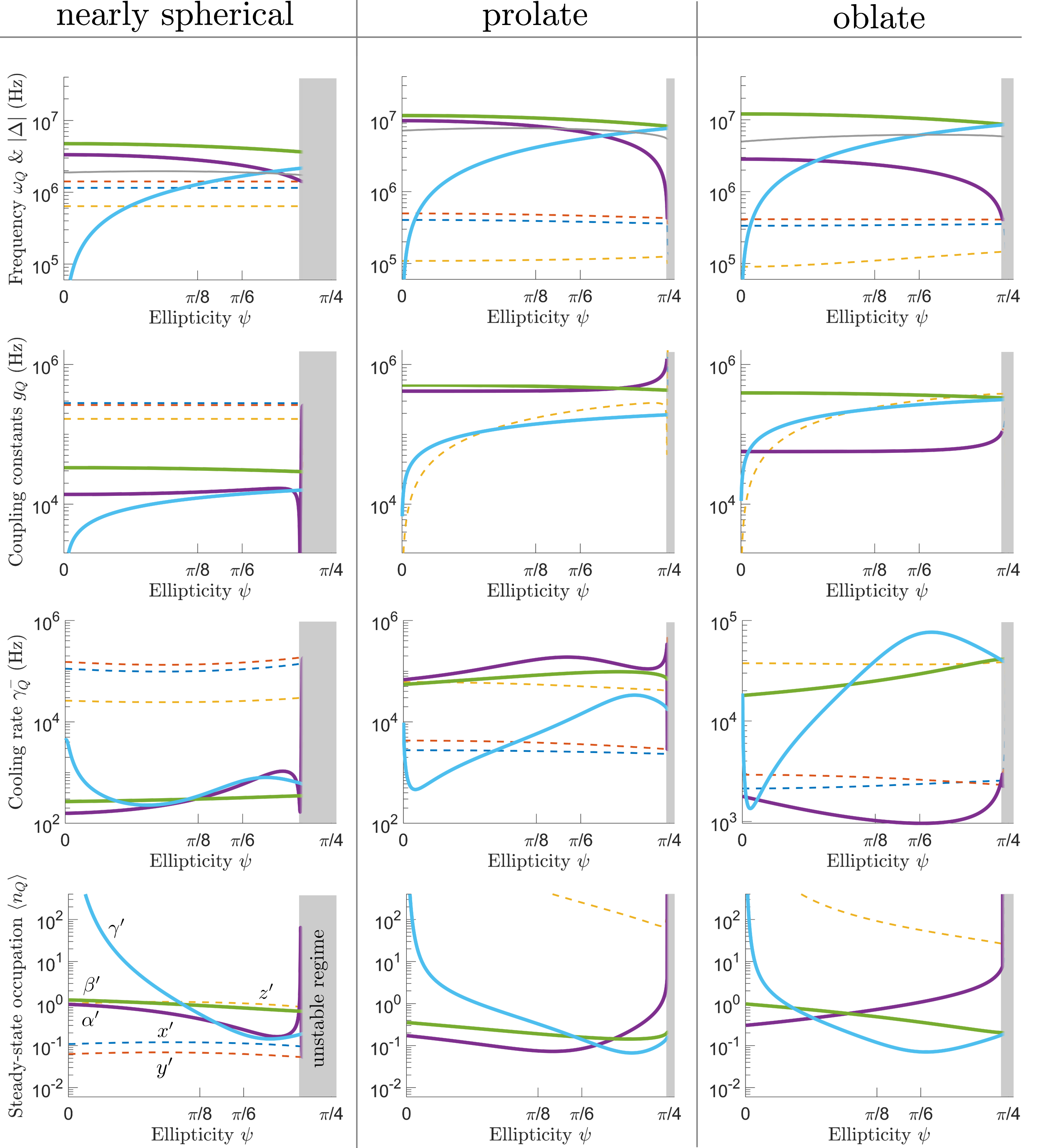}
	\caption{Trapping frequencies, opto-mechanical coupling constants, cooling rates, and steady-state occupations as a function of the tweezer ellipticity for differently shaped particles: a nearly spherical ellipsoid with principal diameters of $(69,70,71)$\,nm, a prolate ellipsoid with $(40,60,140)$\,nm, and an oblate ellipsoid with $(40,90,95)$\,nm. All particles have the same volume. For each particle, the detuning chosen at a given ellipticity is shown in the trapping frequency panel as a thin grey line. Different colors refer to different degrees of freedom: yellow $z'$, dark blue $x'$, red $y'$ (dashed), magenta $\alpha'$, green $\beta'$, light blue $\gamma'$ (solid). Trapping becomes unstable in the shaded area. Cavity parameters for nearly spherical ellipsoid: cavity length $L_{\rm c}=1.5\,$mm, cavity waist $w_c=30\,\mu$m, cavity phase $\phi=3\pi/8$, linewidth $\kappa=600\,$kHz, tweezer power $P=0.1\,$W, tweezer waists $w_x=800\,$nm, $w_y=650\,$nm, tweezer angle $\theta=\pi/4$. Cavity parameters for prolate and oblate ellipsoids:  $L_{\rm c}=3\,$mm,  $w_c=40\,\mu$m, $\phi=0$, $\kappa=2\,$MHz,  $P=0.1$\,W,  $w_x=1.6\,\mu$m, $w_y=1.3\,\mu$m,  $\theta=\pi/2$. We set $\zeta=0$ for simplicity.}
	\label{fig:EllipPro}
\end{figure*}

\section{The steady state}

\subsection{Deep-trapping regime}\label{harmonicpsd}

We now determine the steady state of the master equation \eqref{linmasterequation}, as required for analyzing experiments with deeply trapped nanorotors. We then generalize this treatment to account for the lowest order rotational non-linearities. For this sake, we rewrite the master equation \eqref{linmasterequation} as a set of linearly coupled quantum Langevin equations \cite{bowen2015}. To eliminate the direct mechanical couplings, we diagonalize the  mechanical parts of the Hamiltonians (\ref{linearizedhamiltonians}). This yields two sets of normal modes $Q$, $S_1' = \{x',y',z',\alpha'\}$ and $S_2' = \{\beta',\gamma'\}$ with frequencies $\omega_Q$, coupling constants $g_Q$ and associated heating rates $\xi_Q$ in the rotating wave approximation. These normal mode constants can be calculated from those given in App.~\ref{parameterlist} by numerical diagonalization. For weak cavity fields, $|b_j| \ll \epsilon$, and weak couplings, $|g_{qq'}|^2\ll \omega_q \omega_{q'}$, the normal modes are well approximated by the original modes $S_1$ and $S_2$ \cite{toros2019}.

Defining the Fourier transform of the cavity mode deviations as
\begin{align}\label{Seq:fourierdef}
b_j[\omega] = \frac{1}{\sqrt{2\pi}} \int_{-\infty}^{\infty}\text{d}t \, e^{-i\omega t} \delta b_j(t),
\end{align}
and likewise for the mechanical modes $a_Q$, yields the coupled quantum Langevin equations in Fourier space
\begin{subequations}\label{eq:eom}
	\begin{align}
	-i\omega b_j[\omega] &= (i\Delta_j - \kappa)b_j[\omega] + \sqrt{2\kappa} \, \eta_j [\omega]\nonumber\\ &+ i\sum_{Q\in S_j'} g_Q^* \left(a_Q[\omega] + a_Q^\dagger[-\omega] \right),
	\end{align}
	and
	\begin{align}\label{Seq:mecheq}
	-i\omega a_Q[\omega] &= -i\omega_Q a_Q[\omega] + i\sqrt{\xi_Q} \, \eta_Q[\omega]\nonumber\\ &+ i \left( g_Q b_j[\omega] + g^*_Q b_j^\dagger [-\omega] \right).
	\end{align}
\end{subequations}
The quantum cavity noise operators $\eta_j[\omega]$ account for photon shot noise.  In the time domain, they are characterized by
\begin{subequations}
\begin{align}
\left[ \eta_j(t), \eta_{j'}^\dagger(t') \right] &= \delta_{jj'}\delta(t-t'),\\  \langle \eta_j (t) \eta_{j'} (t') \rangle &= 0, \\
\langle \eta_j^\dagger (t) \eta_{j'} (t') \rangle &= 0, 
%\qquad \langle \eta_Q(t) \eta_{Q'}(t')\rangle = \delta_{QQ'}\delta(t-t')\,,
\end{align}
\end{subequations}
since the thermal  occupation of the cavity can be neglected. Here, $\langle \cdot \rangle$ denotes the ensemble average of the quantum expectation values. The mechanical noise $\eta_Q [\omega]$ accounts for Rayleigh scattering and gas collisions. It is described by classical real white noise, $\langle \eta_Q(t) \eta_{Q'}(t')\rangle = \delta_{QQ'}\delta(t-t')$, implying that $\eta_Q[\omega] = \eta_Q^*[-\omega]$. By writing Eqs.~\eqref{eq:eom}, we assumed that gas collisions only lead to diffusion for the pressures considered here, while damping is negligible.

Solving Eqs. \eqref{eq:eom} for $b_j[\omega]$ yields
\begin{align}\label{Seq:fourierb}
b_j[\omega] = \chi_j[\omega] \left[ \sqrt{2\kappa} \eta_j [\omega] + i\sum_{S_j'} g_Q^* \left(a_Q[\omega] + a_Q^\dagger[-\omega] \right) \right],
\end{align}
with the cavity susceptibility
\begin{equation}
\chi_j[\omega] = \frac{1}{\kappa-i(\Delta_j+\omega)}.    
\end{equation}
Inserting this  into Eq.~(\ref{Seq:mecheq}), neglecting all off-resonant contributions and evaluating $\chi_j[\omega]$ at the mechanical resonance $\omega_Q$ (weak coupling approximation) yield that $a_Q[\omega]$ evolves independently with frequencies $\tilde{\omega}_Q = \omega_Q + |g_Q|^2 \text{Im}(\chi_j[\omega_Q] + \chi_j[-\omega_Q])$ and optomechanical damping rates $\gamma_Q = 2|g_Q|^2 \text{Re}(\chi_j[\omega_Q] - \chi_j[-\omega_Q])$. Solving the resulting equation yields
\begin{align}\label{Seq:fouriera}
& a_Q[\omega] = \chi_Q[\omega]\biggl[ i\sqrt{\xi_Q} \eta_Q [\omega]\nonumber\\ &+ i\sqrt{2\kappa}\left( g_Q \chi_j[\omega]\eta_j[\omega] + g_Q^* \chi_j^*[-\omega] \eta_j^\dagger[-\omega] \right) \biggr],
\end{align}
with the mechanical susceptibilities
\begin{equation}
    \chi_Q[\omega] = \frac{1}{\gamma_Q/2 + i(\tilde\omega_Q - \omega)}.
\end{equation}

The stationary mode operators \eqref{Seq:fourierb} and \eqref{Seq:fouriera} yield the experimentally observable power spectra of the cavity output modes, see below. The stationary mechanical modes can also be used to calculate the stationary mechanical populations,
\begin{equation}
    n_Q(t) = \frac{1}{2\pi} \int {\rm d}\omega {\rm d}\omega' e^{-i (\omega - \omega')t} \langle a^\dagger_Q[\omega] a_Q[\omega']\rangle.
\end{equation}
In the weak-coupling regime, valid for $\gamma_Q^\mp \ll \kappa$ and $\gamma_Q^\mp \gamma_{Q'}^\mp \ll (\omega_Q - \omega_{Q'})^2$ \cite{wilson2008}, with the weak-coupling damping- and heating rates
\begin{equation}
    \gamma_Q^{\mp} = \frac{2|g_Q|^2\kappa}{\kappa^2 + (\Delta_j \pm \omega_Q)^2},
\end{equation}
one obtains the stationary occupation
\begin{equation}\label{eq:ssocc}
    n_Q = \frac{\gamma_Q^+ + \xi_Q}{\gamma_Q^- - \gamma_Q^+}.
\end{equation}
This result also follows from the steady-state condition of the master equation \cite{prlsubmission}.

\begin{figure}[t]
	\centering
	\includegraphics[width=0.8\linewidth]{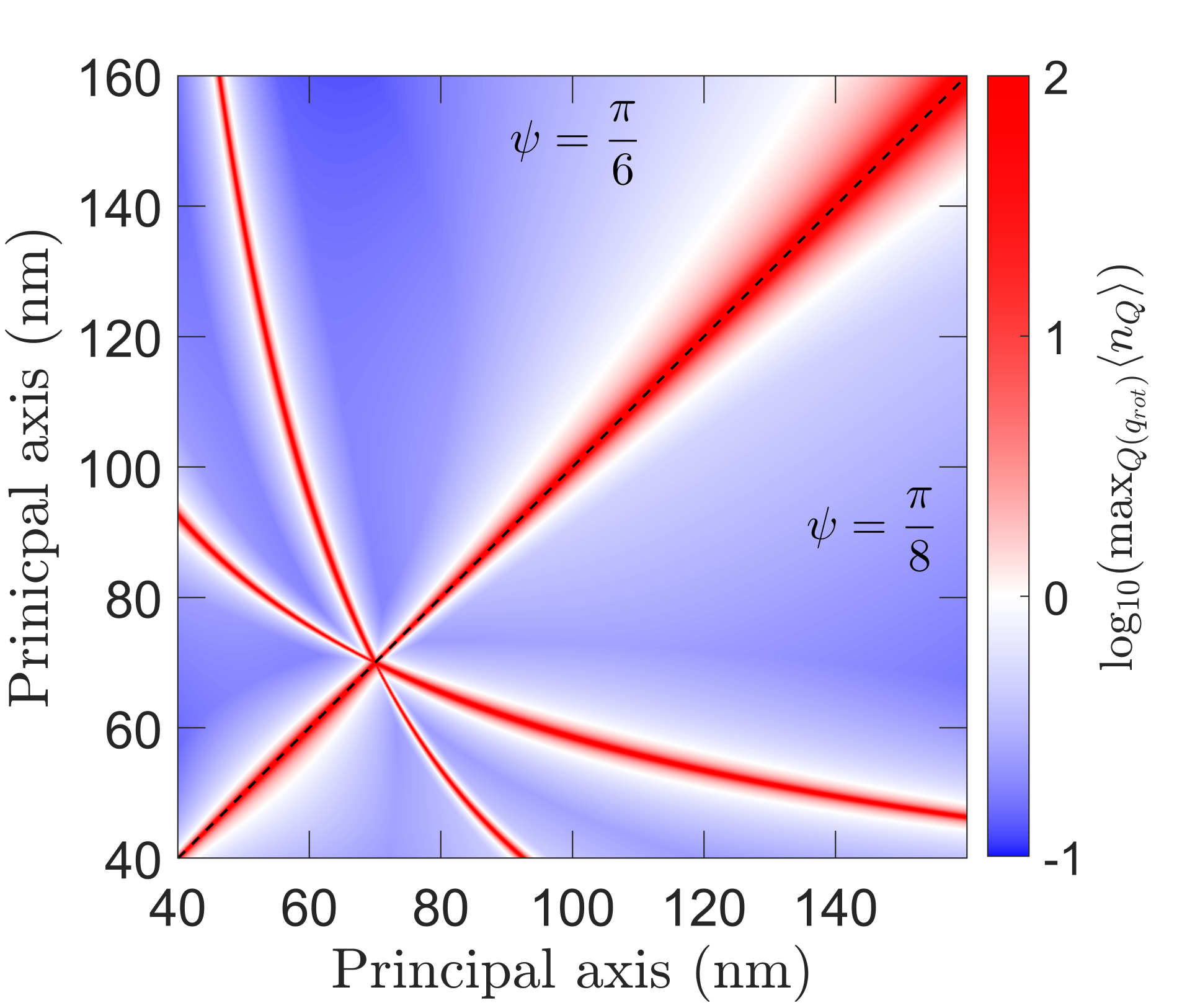}
	\caption{Maximum steady-state librational occupations as a function of the particle shape.
	We vary the length of two principal diameters for an ellipsoid of fixed volume $4\pi (35\,{\rm nm})^3/3$. In the blue region the equilibrium occupation drops below $1$, while red regions correspond to high librational temperatures. For each point, the detuning is chosen as the negative mean of all librational frequencies. All other cavity parameters are as prolate and oblate particles in Fig.\,\ref{fig:EllipPro}.}
	\label{fig:FormRot}
\end{figure}

Figure \ref{fig:EllipPro} shows how the trapping frequencies, opto-mechanical coupling constants, cooling rates, and steady-state occupations depend on the tweezer ellipticity $\psi$ in the deep trapping regime. Three different particle shapes are considered, a nearly spherical ellipsoid, and markedly prolate and oblate ones. The negative detuning is taken as the mean of all librational frequencies for the prolate case and the oblate case, and as the mean of all frequencies for the almost spherical particle. 

We find that $\beta'$ can be cooled for all ellipticities. On the other hand, $\alpha'$ becomes unstable as one approaches circular polarization, $\psi\to\pi/4$, because the radiation pressure torque becomes dominant. Similarly, $\gamma'$ cannot be cooled efficiently as $\psi\to0$ (linear polarization) due to the symmetry of the optical potential. For prolate and almost spherical ellipsoids, efficient librational cooling is found around $\psi=\pi/6$, while for oblate particle shapes cooling around $\psi=\pi/8$ is more efficient. For oblate and prolate particles, librational groundstate cooling and translational groundstate cooling are possible. For nearly spherical particles, it is even possible to simultaneously cool into the combined rotational-translational quantum groundstate \cite{prlsubmission}. The dependence on all other parameters, such as the relative cavity-tweezer alignment $\theta$, largely follows the behavior of spherical particles discussed in Ref. \cite{gonzalez2019}.

Finally, Fig.\,\ref{fig:FormRot} shows the impact of the particle shape on the librational steady-state occupations. Specifically,  the maximum of the three librational occupations is presented as a function of two diameters for an ellipsoid of fixed volume. The tweezer ellipticity is fixed to $\pi/6$ and the detuning is chosen as the negative mean of the librational frequencies. All other parameters are chosen as in Fig. \ref{fig:EllipPro}. Whenever two diameters coincide, librational cooling around the resulting symmetry axis becomes impossible. This can be seen in Fig.\,\ref{fig:FormRot} in the form of three lines of high steady state occupations, intersecting at $(70,70)\,$nm, where the particle is spherical.

\subsection{Rotational nonlinearities}\label{nonlinearpsd}

The nonlinearity in the rotational part of the kinetic energy  $H_0$ gives rise to higher harmonics in the nanoparticle motion. The leading order correction to the harmonic approximation (\ref{eq:h1}) can be calculated by expanding the trigonometric functions in (\ref{eq:h0}) around the equilibrium configuration. Collecting the terms cubic in the mechanical mode operators, one obtains
\begin{align} \label{eq:cubic}
H_{\rm cor} = -\hbar\Bigl[& c_+ (a_\alpha a_\beta a_\gamma - a_\alpha^\dagger a_\beta a_\gamma )\nonumber\\ &+ c_- ( a_\alpha a_\beta^\dagger a_\gamma - a_\alpha a_\beta a_\gamma^\dagger) \Bigr] + \text{h.c.} ,
\end{align}
with $c_\pm = \omega_{\alpha}\alpha_{\rm zp}\beta_{\rm zp}\gamma_{\rm zp} [I_c\omega_{\gamma} \pm (I_a-I_b) \omega_{\beta}]/\hbar$.

While the cubic Hamiltonian \eqref{eq:cubic} only affects the librations, it still couples all normal modes, as follows from diagonalizing the  mechanical part of the Hamiltonian. However, since the normal modes are well approximated by the original modes, this correction is small and we can replace $(\alpha,\beta,\gamma)$ by $(\alpha',\beta',\gamma')$ in Eq.\,\eqref{eq:cubic}.

The resulting equations of motion are equivalent to Eqs.\,\eqref{eq:eom} with an additional term  $i[H_{\rm cor},a_Q]/\hbar$ on the right hand side of \eqref{Seq:mecheq}. We treat this correction as a small perturbation to Eq.\,\eqref{Seq:fouriera} by inserting the stationary solution. Neglecting the cavity vacuum noise in the resulting cubic Hamiltonian, the corrections can be written as
\begin{widetext}
\begin{subequations}\label{corrections}
\begin{align}
a_{\alpha'}^{\rm cor}[\omega] = & -i\sqrt{\frac{\xi_{\beta'}\xi_{\gamma'}}{2\pi}} \chi_{\alpha'}[\omega] \int_{-\infty}^{\infty}\text{d}\omega' \eta_{\beta'}[\omega']\eta_{\gamma'}[\omega-\omega'] \Bigl ( -c_+ \chi_{\beta'}[\omega']\chi_{\gamma'}[\omega-\omega'] + c_+ \chi_{\beta'}^*[-\omega']\chi_{\gamma'}^*[\omega'-\omega]\nonumber\\ 
&+ c_- \chi_{\beta'}^*[-\omega']\chi_{\gamma'}[\omega-\omega']- c_- \chi_{\beta'}[\omega']\chi_{\gamma'}^*[\omega'-\omega] \Bigr),
\end{align}
\begin{align}
a_{\beta'}^{\rm cor}[\omega] = &-i\sqrt{\frac{\xi_{\alpha'}\xi_{\gamma'}}{2\pi}} \chi_{\beta'}[\omega] \int_{-\infty}^{\infty}\text{d}\omega' \eta_{\alpha'}[\omega']\eta_{\gamma'}[\omega-\omega'] \Bigl( c_- \chi_{\alpha'}[\omega']\chi_{\gamma'}[\omega-\omega'] + c_+ \chi_{\alpha'}^*[-\omega']\chi_{\gamma'}^*[\omega'-\omega]\nonumber\\ 
&+ c_- \chi_{\alpha'}^*[-\omega']\chi_{\gamma'}[\omega-\omega']+ c_+ \chi_{\alpha'}[\omega'] \chi_{\gamma'}^*[\omega'-\omega] \Bigr),
\end{align}

\begin{align}
a_{\gamma'}^{\rm cor}[\omega] = &  -i\sqrt{\frac{\xi_{\alpha'}\xi_{\beta'}}{2\pi}} \chi_{\gamma'}[\omega] \int_{-\infty}^{\infty}\text{d}\omega' \eta_{\alpha'}[\omega']\eta_{\beta'}[\omega-\omega'] \Bigl(- c_- \chi_{\alpha'}[\omega']\chi_{\beta'}[\omega-\omega'] + c_+ \chi_{\alpha'}^*[-\omega']\chi_{\beta'}^*[\omega'-\omega]\nonumber\\ &- c_- \chi_{\alpha'}^*[-\omega']\chi_{\beta'}[\omega-\omega'] + c_+ \chi_{\alpha'}[\omega']\chi_{\beta'}^*[\omega'-\omega] \Bigr).
\end{align}
\end{subequations}
\end{widetext}
The convolution originates from Fourier transforming the time-local product of coordinates in the equation of motion.

The non-linear contribtutions to the stationary cavity modes can lead to visible features in the power spectra of the cavity output modes even if they only negligibly affect the steady state occupations \eqref{eq:ssocc}. Since we expect these power spectra to become experimentally relevant, we present a short derivation here.

\subsection{Power spectra}

The power spectral densities (PSD) of the cavity output modes are defined as \cite{bowen2015}
\begin{align}\label{Seq:psd}
S_{b_j^\dagger b_j}[\omega] &= \frac{1}{2\pi}\int_{-\infty}^{\infty}\text{d}\omega' \, \langle b_j^\dagger [\omega] b_j[\omega'] \rangle.
\end{align}
Combining this with Eqs.~(\ref{Seq:fourierb}) and (\ref{Seq:fouriera}) yields the harmonic steady-state power spectra
\begin{widetext}
\begin{align}\label{basepsd}
S^0_{b_j^\dagger b_j}[\omega] = & \frac{1}{2\pi}|\chi_j[\omega]|^2 \Biggl[  2\kappa |\chi_j[-\omega]|^2 \Bigg|\sum_{Q\in S_j'} (g^*_Q)^2 \left( \chi_Q[\omega] - \chi_Q^*[-\omega] \right)\Bigg|^2+ \sum_{Q\in S_j'} |g_Q|^2 \xi_Q | \chi_Q[\omega] - \chi_Q^*[-\omega]|^2 \Biggr].
\end{align}
\end{widetext}
The first term of the PSD originates from cavity vacuum shot noise, which limits cavity cooling even in the absence of mechanical noise. The second factor describes heating due to scattering of tweezer photons and collisions with residual gas atoms.

The cubic corrections can be evaluated by exploiting that 
the unperturbed cavity mode is uncorrelated with its cubic corrections \eqref{corrections}, as are those between the corrections of the different mechanical modes. The perturbed power spectra can thus be written as
\begin{subequations}
\begin{equation}
    S_{b_1^\dagger b_1}[\omega] = S_{b_1^\dagger b_1}^0 [\omega] + S_{b_1^\dagger b_1}^{\alpha'} [\omega],
\end{equation}
and
\begin{equation}
    S_{b_2^\dagger b_2}[\omega] = S_{b_2^\dagger b_2}^0 [\omega] + S_{b_2^\dagger b_2}^{\beta'} [\omega] + S_{b_2^\dagger b_2}^{\gamma'} [\omega]\,.
\end{equation}
\end{subequations}
Keeping only the dominant contributions in the convolution, which result in a Cauchy distribution
\begin{align}
\mathcal{C}_{Q_\pm Q'_\pm}[\omega]= \frac{1}{[(\gamma_Q + \gamma_{Q'})^2/4 + (\omega\mp \tilde\omega_Q \mp \tilde \omega_{Q'})^2]},
\end{align}
yields the corrections
\begin{widetext}
\begin{subequations}
\begin{align}
 S_{b_1^\dagger b_1}^{\alpha'} [\omega] =& 4 \frac{\xi_{\beta'}\xi_{\gamma'}|g_{\alpha'}|^2}{2\pi} \left( \frac{1}{\gamma_{\beta'}} + \frac{1}{\gamma_{\gamma'}} \right) |\chi_1[\omega]|^2 |\chi_{\alpha'}[\omega]-\chi_{\alpha'}^*[-\omega]|^2 \Bigl( c_+^2 \mathcal{C}_{\beta'_+\gamma'_+}[\omega] + c_+^2 \mathcal{C}_{\beta'_-\gamma'_-}[\omega] \nonumber \\
 & + c_-^2 \mathcal{C}_{\beta'_-\gamma'_+}[\omega] + c_-^2 \mathcal{C}_{\beta'_+\gamma'_-}[\omega] \Bigr),
\end{align}
\begin{align}
 S_{b_2^\dagger b_2}^{\beta'} [\omega] = & \frac{\xi_{\alpha'}\xi_{\gamma'}|g_{\beta'}|^2}{2\pi} \left( \frac{1}{\gamma_{\alpha'}} + \frac{1}{\gamma_{\gamma'}} \right) |\chi_2[\omega]|^2  \Bigl ( |c_-\chi_{\beta'}[\omega]-c_+\chi_{\beta'}^*[-\omega]|^2 \Bigl[\mathcal{C}_{\alpha'_+\gamma'_+}[\omega] + \mathcal{C}_{\alpha'_-\gamma'_+}[\omega]\Bigr] \nonumber \\
 &+ |c_+\chi_{\beta'}[\omega]-c_-\chi_{\beta'}^*[-\omega]|^2 \Bigl[ \mathcal{C}_{\alpha'_-\gamma'_-}[\omega] + \mathcal{C}_{\alpha'_+\gamma'_-}[\omega]\Bigr] \Bigr),
\end{align}

\begin{align}
S_{b_2^\dagger b_2}^{\gamma'} [\omega] = & \frac{\xi_{\alpha'}\xi_{\beta'}|g_{\gamma'}|^2}{2\pi} \left( \frac{1}{\gamma_{\alpha'}} + \frac{1}{\gamma_{\beta'}} \right) |\chi_2[\omega]|^2  \Bigl( |c_-\chi_{\gamma'}[\omega]+c_+\chi_{\gamma'}^*[-\omega]|^2 \Bigl[\mathcal{C}_{\alpha'_+\beta'_+}[\omega] + \mathcal{C}_{\alpha'_-\beta'_+}[\omega]\Bigr] + \nonumber \\ 
& +  |c_+\chi_{\gamma'}[\omega]+c_-\chi_{\gamma'}^*[-\omega]|^2 \Bigl[ \mathcal{C}_{\alpha'_-\beta'_-}[\omega] \mathcal{C}_{\alpha'_+\beta'_-}[\omega]\Bigr] \Bigr).
\end{align}
\end{subequations}
\end{widetext}

The peaks due to higher harmonics in these PSDs are genuine signatures of the rotational non-linearities of trapped nanorotors. Such non-linearities could be exploited for quantum protocols and signal transduction applications with deeply trapped rotors. In principle, the influence of higher-order non-linearities can also be calculated  with the strategy outlined here, but at some point the full curvature of the orientation space starts playing a role. This precludes a description in terms of mode operators on the flat tangent space and requires accounting for the full rotation Hamiltonian \eqref{eq:h0} with the quantum potential \eqref{eq:quantpot}. 

\section{Discussion}

In this article, we developed the theoretical framework of elliptic coherent scattering cooling. It provides several theoretical tools, which might become instrumental for future experiments with aspherical nanoparticles. Specifically, we derived a general expression for the non-conservative radiation pressure torque due to scattering of elliptically polarised photons off an ellipsoidally shaped nanoparticle, we presented the corresponding quantum master equation, and we calculated the resulting opto-mechanical trapping and coupling frequencies and studied their dependence on the tweezer ellipticity. Finally, we discussed the signatures of rotational non-linearities in the steady-state power spectra of librationally cooled particles.

While the feasibility and future applications of elliptic coherent scattering cooling are discussed in Ref.~\cite{prlsubmission}, we here  comment briefly on theoretical challenges for future work. A relevant generalization of the theory presented here is to systematically go beyond the Rayleigh-Gans approximation by allowing one of the particle diameters to become comparable with the laser wavelength. This will require adopting the generalized Rayleigh-Gans approximation \cite{schiffer1979,stickler2016} to the coherent scattering setup. Another relevant question is how this cooling scheme can be optimally combined with electric traps and charged particles, as it was successfully achieved with dispersive cavity cooling \cite{millen2015}. This will require taking into account the various electric torques acting on aspherical nanoparticles with non-spherical charge distributions \cite{martinetz2020}. Optimally combining optic with electric techniques might well point the way towards novel trapped quantum interference schemes.

\begin{acknowledgments}
We thank Stephan Troyer for helpful discussions. K.H. acknowledges funding by the Deutsche Forschungsgemeinschaft (DFG, German Research Foundation)--394398290, B.A.S. acknowledges funding from the European Union’s Horizon 2020 research and innovation programme under the Marie Sk\l odowska-Curie grant agreement No. 841040  and by the Deutsche Forschungsgemeinschaft (DFG, German Research Foundation)--439339706. 
H.R. and J.S. contributed equally to this work. 
\end{acknowledgments}
	
\appendix
\begin{widetext}

\section{Euler angles and quantization}\label{app:eulerangles}

The orientation of the particle is specified by Euler angles, defined by subsequent rotations around the $z$-$y'$-$z''$ axes by the angles $\alpha \in (0,2\pi]$, $\beta \in (0,\pi]$, and $\gamma \in (0,2\pi]$, respectively ($z$-$y'$-$z''$ convention) \cite{edmonds1996}. The resulting rotation matrix
\begin{equation}
    {\bf e}_i \cdot R(\Omega) {\bf e}_j =\left [ \left ( \begin{array}{ccc}
    \cos \alpha & -\sin \alpha & 0 \\
    \sin \alpha & \cos \alpha & 0 \\
    0 & 0 & 1
    \end{array} \right )
    \left ( \begin{array}{ccc}
    \cos \beta & 0 & \sin \beta \\
    0 & 1 & 0 \\
    -\sin \beta & 0 & \cos \beta
    \end{array} \right )
    \left ( \begin{array}{ccc}
    \cos \gamma & -\sin \gamma & 0 \\
    \sin \gamma & \cos \gamma & 0 \\
    0 & 0 & 1
    \end{array} \right ) \right]_{ij}
\end{equation}
describes how the body-fixed unit vectors ${\bf n}_k$ can be expressed in the space-fixed coordinate basis, i.e. $\mathbf{n}_k = R(\Omega){\bf e}_k$. Here, matrix elements refer to the space-fixed frame ${\bf e}_k$. 

The canonical angular momenta $p_\Omega$ follow from the classical free rotor Lagrangian as
\begin{subequations}
\begin{align}
    p_\alpha = & \dot{\alpha} \sin^2 \beta \left (I_a \cos^2 \gamma + I_b \sin^2 \gamma+I_c \cot^2 \beta\right )  + \dot{\beta} (I_b - I_a) \sin \beta \sin \gamma \cos \gamma + \dot{\gamma}I_c \cos \beta \\
    p_\beta = & \dot{\alpha} (I_b - I_a) \sin \beta \sin \gamma \cos \gamma + \dot{\beta} (I_a \sin^2 \gamma + I_b \cos^2 \gamma) \\
    p_\gamma = & I_c (\dot{\alpha} \cos \beta + \dot{\gamma}).
\end{align}
\end{subequations}
They are related to the angular momentum vector ${\bf J}$ by
\begin{align}
    p_\alpha = {\bf J}\cdot {\bf e}_z, \qquad p_\beta =  {\bf J}\cdot {\bf e}_\xi, \qquad p_\gamma =  {\bf J} \cdot {\bf n}_3,
\end{align}
where ${\bf e}_\xi = -\sin \alpha\, {\bf e}_x + \cos \alpha\, {\bf e}_y$ is the nodal line of $\beta$ rotations. A straight-forward calculation shows that the free rotor Hamiltonian $H_0 = {\bf J} \cdot I^{-1}(\Omega) {\bf J}/2$ is of the form \eqref{eq:h0}, where $I(\Omega) = \sum_k I_k {\bf n}_k \otimes {\bf n}_k$ is the tensor of inertia.

The rotor can be quantized by promoting the Euler angles and canonical angular momenta to operators \cite{edmonds1996}. In orientation space, the latter take the differential operator form
\begin{equation}
    {p}_\alpha = -i\hbar \partial_\alpha, \qquad p_\beta = -i \hbar \left ( \partial_\beta + \frac{1}{2} \cot\beta \right ), \qquad p_\gamma = -i \hbar \partial_\gamma.
\end{equation}
Note that the canonical momentum operator in $\beta$ contains a contribution due to the curvature of the orientation space. The resulting canonical commutation relations must be formulated in terms of trigonometric functions of the angle operators because only periodic functions are physically admissible. They take the form $[e^{\pm i \mu} , p_\nu] = \mp \hbar \delta_{\mu \nu}e^{\pm i\mu}$ where $\mu,\nu \in \{ \alpha,\beta,\gamma\}$.

The curvature of the configuration space also implies that the orientation-space wavefunction is normalized with respect to the square-root metric determinant, i.e. ${\rm d}\Omega = \sin \beta \,{\rm d}\alpha {\rm d}\beta {\rm d}\gamma$. In addition, when quantizing the Hamiltonian \eqref{eq:h0} one must add the quantum potential \cite{gneiting2013}
\begin{equation} \label{eq:quantpot}
    Q(\Omega) = -\frac{\hbar^2}{16} \left ( \frac{1}{I_a} + \frac{1}{I_b} \right ) \left (\frac{1}{\sin^2 \beta} +1 \right ) + \frac{\hbar^2}{16}  \left ( \frac{1}{I_a} - \frac{1}{I_b} \right ) \left ( \frac{5}{\sin^2 \beta} -3 \right ) \cos 2\gamma .
\end{equation}

\section{Quadratic correction to the electric field integral equation}\label{app:d}

The electric field integral equation~(\ref{eq:3}) including the $k^2$-corrections reads as
\begin{align}
	\mathbf{E}_{\rm cor}(\mathbf{r}) \simeq  \left (\frac{k^2}{8\pi} C({\bf r}) + \frac{iVk^3}{6\pi}\right )  \chi\mathbf{E}(\mathbf{r}) + \frac{\varepsilon - 1}{4\pi} \int_{V({\bf R},\Omega)} \text{d}^3 r'  \left [ 3 (\mathbf{r}-\mathbf{r}')\otimes (\mathbf{r}-\mathbf{r}') - \mathds{1}|\mathbf{r}-\mathbf{r}'|^2 \right ] \frac{\mathbf{E}_{\rm cor}(\mathbf{r}')}{|\mathbf{r}-\mathbf{r}'|^5}.
\end{align}
with the shape-dependent correction tensor
\begin{equation}
    C({\bf r})=\int_{V({\bf R},\Omega)} \frac{\text{d}^3 r'}{|\mathbf{r}-\mathbf{r}'|} \left [\frac{\mathbf{r}-\mathbf{r}'}{|\mathbf{r}-\mathbf{r}'|} \otimes \frac{\mathbf{r}-\mathbf{r}'}{|\mathbf{r}-\mathbf{r}'|} + \mathds{1} \right ].
\end{equation}
It is an integral  equation for $\mathbf{E}_{\rm cor}$, where  $\mathbf{E}$ is taken as constant inside the particle. In contrast, the shape-dependent correction tensor varies inside the particle, implying that the corresponding equation cannot be solved straight-forwardly. However, since this term is real it only contributes to the susceptibility on the order $k^2 V^{2/3}$, which is negligible for the particles considered in this paper.

\section{Phase-space contraction rate}\label{phasespace}
		
To quantify whether non-conservative generalized forces lead to a local phase-space contraction or expansion, we consider the dynamics of the $2N$-dimensional phase-space point ${\bf z}_t = (q_1,\ldots,q_N,p_1,\ldots,p_N)$, where $N$ denotes the number of degrees of freedom. Including non-conservative generalized contributions, we can write the equations of motion in the form $\dot{\bf z}_t = J \partial_{\bf z} H + {\bf K}$, where $J$ is the $(2N \times 2N)$-symplectic matrix, $J \partial_{\bf z} = (\partial_{p_1}, \ldots,\partial_{p_N},-\partial_{q_1},\ldots, -\partial_{q_N})$, $H \equiv H({\bf z}_t)$ is the Hamiltonian and the vector field ${\bf K} \equiv {\bf K}({\bf z}_t)$ describes the non-conservative part of the dynamics.
		
For non-vanishing ${\bf K}$, the infinitesimal volume element ${\rm d}V (\mathbf{z}_t) = {\rm d}^{2N} {\bf z}_t$ is in general a function of time $t$. To quantify the resulting contraction or expansion rate, we introduce the dynamical mapping $\Phi_t$ from an arbitrary phase space point ${\bf z}_0$ to its time-evolved coordinates ${\bf z}_t = \Phi_t({\bf z}_0)$. Thus, we can write ${\rm d}V(\mathbf{z}_t) = \Psi_t(\mathbf{z}_0) \text{d}V(\mathbf{z}_0)$, where $\Psi_t(\mathbf{z}_0) = \det[\partial_{0} \otimes \Phi_t (\mathbf{z}_0)]$ is the Jacobi-determinant of the dynamical mapping $\Phi_t$, and $\partial_0$ denotes the derivative with respect to the initial conditions ${\bf z}_0$.
		
The time derivative of the Jacobi determinant can be calculated by differentiating the determinant function,
\begin{align} \label{eq:psidot}
		\partial_t \Psi_t(\mathbf{z}_0) = \text{tr}\left[  [\partial_{0} \otimes \Phi_t ( \mathbf{z}_0)]^{-1} \partial_t [\partial_{0} \otimes \Phi_t (\mathbf{z}_0) ] \right]\, \Psi_t(\mathbf{z}_0).
\end{align}
Thus each infinitesimal volume element $dV({\bf z}_t)$ locally contracts or expands with a rate quantified by the right-hand side of Eq.~\eqref{eq:psidot}. Using the equations of motion and applying the chain rule, e.g. $\partial_0 \otimes {\bf K}({\bf z}_t) = [\partial_0 \otimes \Phi_t({\bf z}_0)] [ \partial \otimes {\bf K}({\bf z}_t) ]$ yields the local phase space contraction rate
\begin{align}\label{eq:psrate}
		\Gamma_{\rm c}({\bf z}_t) = -\partial_{\bf z} \cdot \mathbf{K}(\mathbf{z}_t).
\end{align}
If the non-conservative contribution only acts as a force $F_q$, the contraction rate takes the form of a momentum divergence, $\Gamma_{\rm c} = -\sum_{q} \partial_{p_q} F_q$.
		
\section{Deep-trapping parameters}\label{parameterlist}
		
We provide the explicit expressions for  all harmonic frequencies and coupling constants arising in the deep-trapping regime. Writing the frequencies as $\omega_q^2 = -2\hbar U_0 v_q$, we have
\begin{subequations}
\begin{align}
		v_x =& \frac{1}{m w_x^2}\Bigl[ 2\epsilon^2(\chi_c \cos^2 \psi + \chi_b\sin^2\psi)+ \epsilon \cos\phi (k^2 w_x^2\sin^2\theta + 2)\text{Re}[b_0^*(\chi_c\cos(\theta-\zeta)\cos\psi-i\chi_b\sin(\theta-\zeta)\sin\psi)] \nonumber \\ 
		&+ |b_0|^2 k^2 w_x^2\cos 2\phi \sin^2\theta(\chi_c \cos^2 (\theta-\zeta) + \chi_b\sin^2(\theta-\zeta))\Bigr], \\
		v_y =& \frac{1}{m w_y^2}\Bigl[ 2\epsilon^2(\chi_c \cos^2 \psi + \chi_b\sin^2\psi)+ \epsilon \cos\phi (k^2 w_y^2\cos^2\theta + 2)\text{Re}[b_0^*(\chi_c\cos(\theta-\zeta)cos\psi-i\chi_b\sin(\theta-\zeta)\sin\psi)] \nonumber \\
		&+ |b_0|^2 k^2 w_y^2\cos 2\phi \cos^2\theta(\chi_c \cos^2 (\theta-\zeta) + \chi_b\sin^2(\theta-\zeta))\Bigr], \\
		v_z = & \frac{1}{m z_{\rm R}^2}\Bigl[ \epsilon^2(\chi_c \cos^2 \psi + \chi_b\sin^2\psi) + \epsilon \cos\phi [1+(z_{\rm R}k-1)^2]\text{Re}[b_0^*(\chi_c\cos(\theta-\zeta)\cos\psi-i\chi_b\sin(\theta-\zeta)\sin\psi)]\Bigr], \\
		v_{\alpha} = & \frac{\chi_c-\chi_b}{I_a} \Bigl[ \epsilon^2\cos 2\psi + 2\epsilon \cos\phi\text{Re}[b_0^*(\cos(\theta-\zeta)\cos\psi+i\sin(\theta-\zeta)\sin\psi)] + |b_0|^2 \cos^2\phi \cos (2\theta-2\zeta) \Bigr],\\
		v_{\beta} = & \frac{\chi_c-\chi_a}{I_b} \Bigl[ \epsilon^2\cos^2\psi + 2\epsilon \cos\phi \cos(\theta-\zeta)\cos\psi\text{Re}[b_0] + |b_0|^2 \cos^2\phi \cos^2(\theta-\zeta) \Bigr],\\
		v_{\gamma} = & \frac{\chi_b-\chi_a}{I_c} \Bigl[ \epsilon^2\sin^2\psi - 2\epsilon \cos\phi \sin(\theta-\zeta)\sin\psi\text{Im}[b_0] + |b_0|^2 \cos^2\phi \sin(2\theta-2\zeta) \Bigr],
\end{align}
\end{subequations}
with the tweezer-minimum amplitude of mode $1$ given by
\begin{equation}
    b_0 = [b_{\rm s}(q_{\rm tw})]_1 =  -\frac{iU_0\epsilon}{\kappa-i\Delta_1} \cos\phi [\chi_c\cos(\theta-\zeta)\cos\psi - i\chi_b \sin(\theta-\zeta)\sin\psi].
\end{equation}
The detunings in the harmonic regime are given by
\begin{subequations}
\begin{align}
\Delta_1 &= [\Delta_{\rm eff}(q_{\rm tw})]_{11} = \Delta - U_0 (\chi_c \cos^2\theta + \chi_b\sin^2\theta) \cos^2\phi,\\
\Delta_2 &= [\Delta_{\rm eff}(q_{\rm tw})]_{22} = \Delta - U_0 \chi_a \cos^2\phi.
\end{align}
\end{subequations}

The mechanical-mechanical couplings are symmetric, $g_{qq'} = g_{q'q}$, and take the form $g_{qq'} = U_0 q_{\rm zp} q_{\rm zp}' G_{qq'}/2$ with
\begin{subequations}
\begin{align}
		G_{xy} = & \frac 1 2 k^2 \sin 2\theta \cos\phi \Bigl[  \epsilon\text{Re}[b_0^* (\chi_c \cos(\theta-\zeta)\cos\psi-i\sin(\theta-\zeta)\sin\psi)] \nonumber\\ &+ 2 |b_0|^2 \sin\phi (\chi_c \cos^2(\theta-\zeta) + \chi_b \sin^2(\theta-\zeta)) \Bigr], \\
		G_{xz} = & - k^2 \epsilon \left( 1-\frac{1}{k z_{\rm R}} \right) \sin\phi\sin\theta\text{Im}[b_0^* (\chi_c\cos(\theta-\zeta)\cos\psi -i\chi_b\sin(\theta-\zeta)\sin\psi)], \\
		G_{yz} = & G_{xz} \cot\theta, \\
		G_{x\alpha} = & - k (\chi_c-\chi_b) \sin\theta \sin\phi \Bigl[ \epsilon \text{Re}[b_0^* (\sin(\theta-\zeta)\cos\psi-i\cos(\theta-\zeta)\sin\psi)] + |b_0|^2 \cos\phi \sin(2\theta-2\zeta) \Bigr], \\
		G_{y\alpha} = & G_{x\alpha}\cot\theta,\\
		G_{z\alpha} = & - k \epsilon (\chi_c-\chi_b) \left( 1 - \frac{1}{k z_{\rm R}} \right) \cos\phi\text{Im} [b_0^* (\sin(\theta-\zeta)\cos\psi - i\cos(\theta-\zeta)\sin\psi)], \\
		G_{\beta\gamma} = & \frac 1 2 (\chi_b-\chi_a) \cos\phi \Bigl[  2\epsilon \text{Re}[b_0^* (\sin(\theta-\zeta)\cos\psi -i\cos(\theta-\zeta)\sin\psi)] + |b_0|^2 \cos\phi \sin(2\theta-2\zeta) \Bigr].
\end{align}
\end{subequations}
All other couplings vanish.
		
Finally, the opto-mechanical couplings read as $g_{jq} = U_0 q_{\rm zp} G_{jq}$ with
\begin{subequations}
\begin{align}
		G_{1x} =& k \sin\theta \sin\phi \Bigl[ \epsilon (\chi_c\cos(\theta-\zeta)\cos\psi+i\chi_b\sin(\theta-\zeta)\sin\psi) + 2 b_0^* \cos\phi (\chi_c\cos^2(\theta-\zeta) +\chi_b\sin^2(\theta-\zeta))\Bigr], \\
		G_{1y} = & G_{1x} \cot\theta , \\
		G_{1z} = & i k \epsilon \left( 1-\frac{1}{k z_{\rm R}} \right) \cos\phi (\chi_c\cos(\theta-\zeta)\cos\psi+i\chi_b\sin(\theta-\zeta)\sin\psi),\\
		G_{1\alpha} = & (\chi_c-\chi_b) \cos\phi \Bigl[ \epsilon (\sin(\theta-\zeta)\cos\psi + i \cos(\theta-\zeta)\sin\psi) +b_0^* \cos\phi\sin(2\theta-2\zeta) \Bigr],\\
		G_{2\beta} = & (\chi_c-\chi_a) \cos\phi (\epsilon\cos\psi + b_0^* \cos(\theta-\zeta)\cos\phi),\\
		G_{2\gamma} = & (\chi_b-\chi_a) \cos\phi (i\epsilon\sin\psi + b_0^* \sin(\theta-\zeta)\cos\phi).
\end{align}
\end{subequations}

\end{widetext}

\end{document}